\documentclass[preprint,showkeys,amsmath,amssymb,superscriptaddress]{revtex4}
\usepackage[dvips]{graphicx}% Include figure files
\usepackage{enumitem}
\usepackage{dcolumn}% Align table columns on decimal point
\usepackage{bm}% bold math
\usepackage[titletoc,title]{appendix}
\usepackage{longtable}
\usepackage{bm}
\usepackage{color}

\begin{document}
\setlength\LTcapwidth{6.5in}
\setlength\LTleft{0pt}
\setlength\LTright{0pt}

\title[Critical temperature determination]{Critical temperature determination on a square-well 
fluid using an adaptation of the Microcanonical-ensemble computer simulation method.}

\author{Francisco Sastre}
\email{sastre@fisica.ugto.mx}
\affiliation{%
Departamento de Ingenier\'ia F\'isica,\ Divisi\'on de Ciencias e Ingenier\'ias,\\
Campus Le\'on de la Universidad de Guanajuato%
}

\date{\today}% It is always \today, today,

\begin{abstract}
In this work we present a novel method to evaluate the liquid-vapor critical temperature using a generalization of the Microcanonical-ensemble computer simulation method (MCE). The isotherms of the chemical potential
versus densities are obtained for a square-well (SW) fluid with interaction range
$\lambda/\sigma = 1.5$: From these curves we extracted the critical temperature for different system sizes observing the change of the slope on the curves of the chemical potential in the critical region as function of the temperature.
Working with different systems sizes and Finite Size Scaling (FSS) Theory the critical temperature $T_c=1.2180(29)$ and the critical exponent $\nu=0.65(3)$ are obtained, without previous knowledge of $T_c$ or $\nu$. These results are in good agreement with the reported values for this system.
\end{abstract}

%\pacs{Valid PACS appear here}


\maketitle

\section{Introduction}

%\vspace{0.2cm}
Computer simulations are a common tool to study the critical behavior in fluids.  Various programming
algorithms and techniques have been developed in order to study the phase boundaries, critical points and the
universality classes of fluids with different interaction
potentials~\cite{Panagiotopoulos1987,Ferremberg1988,Demiguel1997,Orkoulas1999,Kim2005,Kofke2006,Zhao2017}.
Numerical simulations are restricted to finite systems, nevertheless the finite size scaling (FSS) theory~\cite{Fisher1972}
allows us to extrapolate the results obtained from finite systems in order to extract the critical properties of the infinite system.
For the evaluation of the critical point in fluids we need to evaluate two separate parameters, the critical
temperature and the critical density. This is due to a lack of a well defined axis of symmetry, unlike magnetic
materials and other condensed matter systems where the evaluation of the critical point requires just one
parameter, generally the critical temperature. The aim of this work is to show that it is possible to evaluate
just one parameter, the critical temperature, at least in a system with a moderate asymmetry, using a
novel methodology that requires the calculated values of the chemical potential. Where the chemical potential
curves as function of the density are obtained with an efficient
algorithm derived from the Microcanonical-ensemble computer simulation method (MCE)~\cite{Sastre2015,Sastre2018}.

We applied the method in the
square-well (SW) fluid system, that is considered as the simplest non-trivial model that capture the main
phenomenology of real atomic fluids~\cite{Barker1976}. The SW potential incorporates a hard sphere repulsion and a
finite range attraction. The SW particles of diameter $\sigma$ interact with the potential
\begin{equation}
\phi(r)=\left\{
\begin{array}{ccc}
\infty & \mbox{if} & r\leq \sigma \\
-\epsilon & \mbox{if} & \sigma < r \leq \lambda\sigma \\
0 & \mbox{if} &  r > \lambda\sigma \\
\end{array}
\right.,\label{potencial}
\end{equation}
where $\epsilon$ is the depth-well energy and $\lambda$ is the range of the attractive interaction. 
When $\lambda=1.5$ the SW fluid  
has as advantage that presents a not so strong asymmetry in the Vapor Liquid Equilibrium (VLE) phase diagram in
the vicinity of the critical region. This feature seems to be associated to a small value of the Yang-Yang ratio
$ R_\mu= -0.08(12)$~\cite{Orkoulas2001}, in contrast to the strong asymmetry of the Restrict Primitive Model (RPM) whose ratio is $R_\mu \simeq 0.26$~\cite{Kim2003}.

In the next section the numerical algorithm used in this work will be explained. In Section~\ref{resultados} the results and
the simulation details for the SW fluid system are presented. The concluding remarks close the paper in Section~\ref{conclusiones}.

\section{Simulation Method}

As a first step we will explain the basic points of the MCE method, the complete explanation can be found on
Ref.~\cite{Sastre2015}. The MCE method allows the direct
evaluation of the inverse temperature as function of the internal energy using the microcanonical relation
\begin{equation}
\frac{P^{(m)}_{\nu\mu}}{P^{(m)}_{\mu\nu}} =\frac{\Omega(E_\mu)}{\Omega(E_\nu)}, 
\end{equation}
where $P^{(m)}_{\nu\mu}$ is the probability, in the microcanonical ensemble, to reach the macrostate $\Omega(E_\mu)$ starting from the macrostate
$\Omega(E_\nu)$ and $P^{(m)}_{\mu\nu}$ is the reversal probability, and
$\Delta E = E_\mu - E_\nu = \eta\epsilon$, with
$\eta$ integer. This fact is used to obtain the inverse temperature,
\begin{equation}
\Delta S = k_B (\ln{P^{(m)}_{\nu\mu}}-\ln{P^{(m)}_{\mu\nu}}) \approx \eta\epsilon \frac{1}{T},
\end{equation}
with $N$ and $V$ fixed. The algorithm works using random displacement of particles to take the system to the
different energy levels allowed in the simulation, {\it i.e.} the particle displacement is the mechanism that
generate new microstates. 

In this work we will be focused in the evaluation of the chemical potential, thus the fixed $N$ condition must be
removed. The system can reach new macrostates inserting or removing particles at random, this is the new
mechanism, and now
$P^{(m)}_{(\nu j)(\mu i)}$ will be the probability to reach the macrostate with energy $E_\mu$ and number of particles
$N_i$ starting from the macrostate with energy $E_\nu$ and number of particles $N_j$. 
In this case the change on the entropy 
will be
\begin{equation}
\Delta S = k_B \ln{(P^{(m)}_{(\nu j)(\mu i)}/P^{(m)}_{(\mu i)(\nu j)})}
\approx \Delta E\frac{1}{T} + \gamma\frac{\mu}{T},
\label{chemical1}
\end{equation}
where $\gamma = \pm 1 = \Delta N$. The last equation depends on $E$ and $N$, where the parameters $\mu/T$ and $1/T$
can be extracted from the simulations. At this point, if we want to perform the simulations in the microcanonical ensemble, it will be necessary to
keep the record of the changes in the energy and the number of particles in the system. In order to avoid the dependence on $E$, {\it i.e.}
just keep track of the changes in the number of particles, a heat-bath is incorporated to the
simulation at a fixed $T$. The probabilities are now $P\propto\Omega(E,N) e^{-E/k_B T}$, where the absence of the superscript indicates that 
we are no longer in the microcanonical ensemble, and the
ratio of the probabilities will be given by
\begin{equation}
\frac{P_{(\nu j)(\mu i)}}{P_{(\mu i)(\nu j)}} = \frac{P^{(m)}_{(\nu j)(\mu i)}}{P^{(m)}_{(\mu i)(\nu j)}} e^{-(E_\nu -E_\mu)/k_B T}
=\frac{\Omega(E_\mu,N_i)}{\Omega(E_\nu,N_j)} e^{\Delta E/k_B T}.
\label{chemical2}
\end{equation}
Combining Eqs. (\ref{chemical1}) and (\ref{chemical2}) the chemical potential can be obtained with
\begin{equation}
\ln{(P_{(\nu j)(\mu i)}/P_{(\mu i)(\nu j)})}
\approx \gamma\frac{\mu}{kT}.
\label{chemical3}
\end{equation}
As the right hand side of the last equation is now $\Delta E$ independent we can drop the subscripts $\mu$ and $\nu$
in the probabilities.
In the simulation the probabilities now can be estimated with the rate of attempts $T_{ji}$ to go from a macrostate
with $N_j$ particles (level $j$) to a macrostate with $N_i$ (level $i$). The quantity $T_{ji}$ is given by
the relation
\begin{equation}
T_{ji} = \frac{z_{ji}}{z_j},
\label{ratios}
\end{equation}
where $z_{ji}$ is the number of times that the system attempts to change from level $j$ to level $i$ and $z_j$
is the number of times that the system spends in level $j$. For the estimation of the $z_{ji}$ and $z_i$ values,
the detailed steps are:
\begin{enumerate}[label=(\roman*)]
\item With $j$ as the initial state, we choose a point coordinate within the simulation box at random
and $z_j$ is always incremented by 1.
\item If a particle is located at the chosen coordinate its removal would lead to a state with $N_i=N_j-1$,
otherwise an insertion of a particle centered at the coordinate would lead to the state $N_i = N_j+1$.
\item If $N_i$ is an allowed particle level we evaluate $\Delta E$ between states $N_i$ and $N_j$, and
the quantity $z_{ji}$ is incremented by 1 with probability min$(1,e^{-\Delta E/kT})$.
\item The particle remove/insertion attempt is accepted with probability min$(1,\frac{z_{ij}z_j}{z_{ji}z_i})$.
This condition assures that all levels are visited with equal probability, independently of their degeneracy.
\end{enumerate}
The values $z_j$ and $z_{ji}$ are initialized to 1 and after a large number of particle remove/insertion
attempts we will observe that $T_{ji}\to P_{j i}$.
The main difference with respect to the original MCE algorithm is step (iii), where the heat-bath
condition is incorporated.

Once that the quantities $z_{ji}$ and $z_i$ are obtained the chemical potential can be evaluated with the
relation
\begin{equation}
\mu^* (\rho^*_i) = \frac{1}{2}\left[(\ln{(T_{i,i-1}/T_{ i-1, i})}-\ln{(T_{i,i+1}/T_{ i+1, i})}\right],
\label{chemical4}
\end{equation}
where we used the reduced units $V^*=1$, $\rho^*_i=\sigma^3 \frac{N_i}{V}=\sigma^3 N_i$, $T^*=kT/\epsilon$ and $\mu^*=\mu/T^*$.

An additional advantage of the algorithm is that it is possible to restrict 
the particle levels discarding all cases where $N_i < N_{\mbox{\tiny min}}$ or $N_i > N_{\mbox{\tiny max}}$.

\begin{figure}[!h]
\begin{center}
\includegraphics[width= 10.0cm]{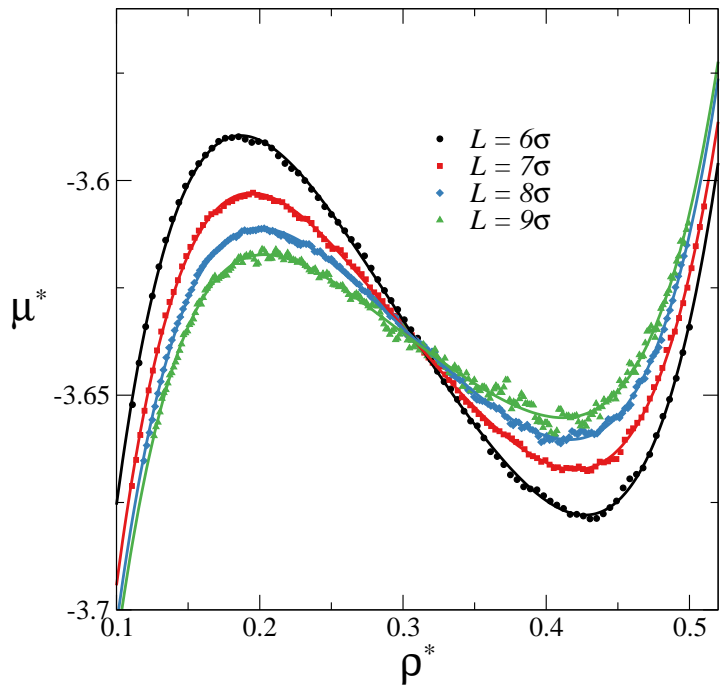}
\caption{\label{isotermas} 
Reduced chemical potential vs. reduced density for $T^*=1.208$, $\lambda=1.5$ and simulation box 
length $L/\sigma=6,~7,~8$
and 9, from top to bottom on the left side of the graph. The symbols are the simulation results and the solid
lines are fifth order polynomial fits to simulation data.
}
\end{center}
\end{figure}

The simulations were carried out in an unitary box with periodic boundary conditions, and different system
sizes. The input values are the reduced temperature and the reduced density intervals, where 
the simulations will be restricted. As an
example for $L=8\sigma$ ($\sigma=0.125$) and $0.10\le\rho^*\le 0.50$ the simulation is restricted to $N_{\mbox{\tiny
min}}=51$ and $N_{\mbox{\tiny max}}=256$ particles.
In Fig.~\ref{isotermas} we are showing the isotherm $T^* = 1.208$ for four different system sizes $L/\sigma
= 6,~7,~8$ and 9. The curves were obtained using up to $1.5 N_{\mbox{\tiny max}}\times 10^7$
particle removal/insertion attempts and four different independent runs for every set of parameters. 

\section{Results}
\label{resultados}

\begin{figure}[!h]
\begin{center}
\includegraphics[width= 10.0cm]{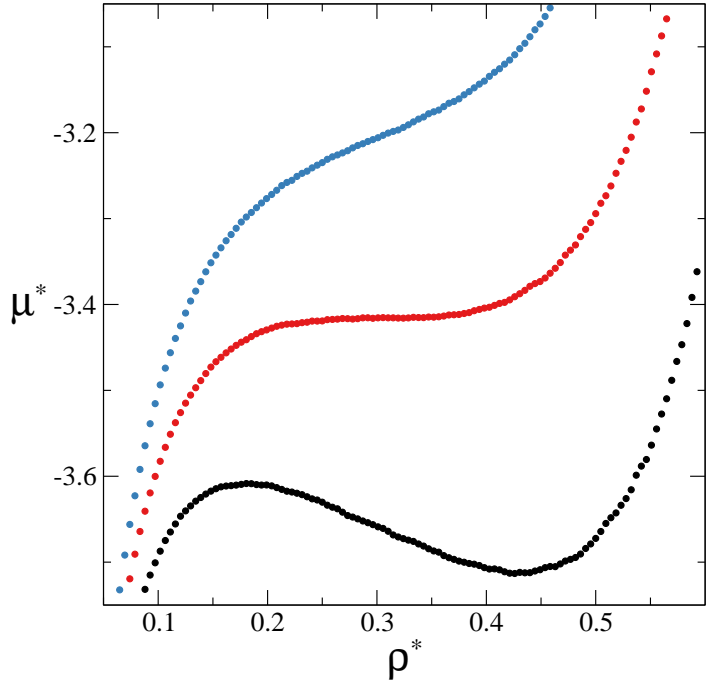}
\caption{\label{pendientes} 
Reduced chemical potential vs. reduced density for a system with size $L=6\sigma$ and $T^*=1.36,~1.28$ and
$1.20$, from top to bottom. We can observe that the simulation is able to capture a clear change of slope 
around $\rho^*=0.3$. 
}
\end{center}
\end{figure}

We can observe that the method is able to capture the behavior of the chemical potential in the
coexistence region, where the unstable phase, $\partial\mu^*/\partial\rho^* < 0$, is clearly visible.
It is possible to obtain the coexistence densities from these curves 
using the equal area rule, that 
can be derived from the Gibbs-Duhem equation and the equilibrium
conditions for pressures $P^*_v=P^*_l$ and chemical potentials $\mu^*_v=\mu^*_l$, where the subscripts $v$ and
$l$ indicates the vapor and liquid phase respectively. Table~\ref{temperatura1208} shows the
estimation of the coexistence densities and the equilibrium chemical potential, $\mu^*_{\mbox{\tiny eq}}$,
obtained from the data shown in Fig.~\ref{isotermas}.
\begin{center}
\begin{table}[ht!]
\caption{\label{temperatura1208}
Vapor-liquid coexistence and equilibrium chemical potential data for $\lambda=1.5$ and $T^*=1.208$ for four
system sizes. The values between parenthesis indicate
the uncertainty in the last digits.
}
\begin{tabular}{ccccccc}
\hline
\hline
$L/\sigma$ & ~ & $\rho^*_v$ & ~ & $\rho^*_l$ & ~ & $\mu^*_{\mbox{\tiny eq}}$\\
\hline
  6 & ~ & 0.1197(3)  & ~ & 0.4988(20) & ~ & $-3.6350(5)$ \\ 
  7 & ~ & 0.1312(2)  & ~ & 0.4887(6) & ~ & $-3.6363(3)$ \\ 
  8 & ~ & 0.1400(8) & ~ & 0.4798(25) & ~ & $-3.6364(4)$ \\ 
  9 & ~ & 0.1465(21) & ~ & 0.4724(24) & ~ & $-3.6369(3)$ \\ 
\hline
\hline
\end{tabular}
\end{table}
\end{center}
It must be point out that the chemical potentials curves obtained here seem to be shifted with respect to those reported
by Del R\'io {\em et al.}~\cite{Delrio2002}, but this fact does not affect the estimated coexistence densities.
It is possible then to obtain the VLE curve using this method, and from this curve the critical values for the
temperature and the density. This evaluation will be left for future works and here another approach will be explored.
In the supercritical phase the negative slope must not be present in the chemical potential curves, so for every
systems size it should exist a ``critical temperature'', $T^*_c(L)$, where the slope around the critical density
changes its sign. Fig.~\ref{pendientes} presents the simulation data for three different temperatures:
above, around and below the critical temperature for $L=6$, $T^*_c(6)$. Since the curves are fairly linear around $\rho^*\simeq 0.3$,
and as the SW with $\lambda=1.5$ is fairly symmetric, it will be possible to evaluate the temperature value where
the change of slope is
located with a relatively small computational effort. The simulations can be restricted to a very narrow number
of particle number levels in order to evaluate with high accuracy the slopes of the chemical potential.

\begin{figure}[!h]
\begin{center}
\includegraphics[width= 10.0cm]{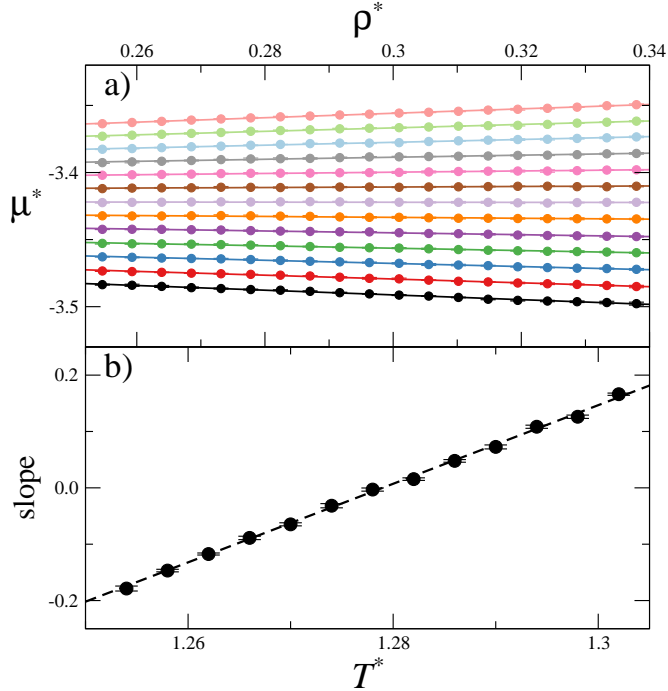}
\caption{\label{escalamiento} 
a) Chemical potential as function of the density for $L=6\sigma$ with temperatures in the interval
$1.254\le T^* \le 1.302$ with $\Delta T^* = 4\times 10^{-3}$, decreasing from
top to bottom. Solid lines are linear fits to the simulation data. b) Slopes from the linear fits of a) as
function of the temperature. With a linear fit we obtained the value $T^*_c(6)=1.27898(20)$. 
}
\end{center}
\end{figure}

For the evaluation of the critical temperature we performed simulations with system sizes $L/\sigma =6.0,~6.5,~7.0,~7.5,~8.0$ and 9.0 and up to $2 N_{\mbox{\tiny max}}\times 10^8$ particle removal/insertion attempts, using four different independent runs for every set of parameters. 
In Fig.~\ref{escalamiento}.a we are showing the results for a system of size $L=6\sigma$ for several temperatures, higher
temperatures are in the top of the graphs. In this region it is possible to estimate the slope
with linear fits to the simulation data. In Fig.~\ref{escalamiento}.b we present the slopes as function of $T*$. 
Using the same procedure for each one of the system sizes considered in this work the curve of
$T^*_c$ as function of $L$ can be obtained. From here a non-linear curve fit is performed to the scaling relation
\begin{equation}
T^*_c(L) = T^*_c+a L^{-1/\nu},
\label{escala}
\end{equation}
where $T^*_c$ is the critical temperature in the thermodynamic limit, $a$ is a non universal parameter and $\nu$ is the correlation length critical
exponent.
 
\begin{figure}[!h]
\begin{center}
\includegraphics[width= 10.0cm]{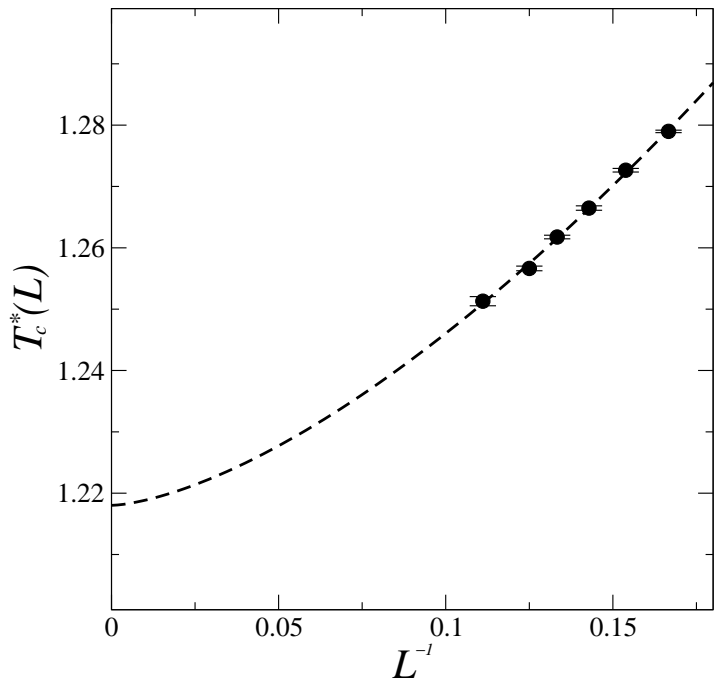}
\caption{\label{critico} 
Evaluation of the critical temperature and the correlation length critical exponents. Black dots are the results
according to the simulation method presented in this work and the dashed line is a non-linear curve fit to Eq.~(\ref{escala}).
The results from the fit are $T^*_c=1.2180$, $\nu=0.65$ and $a=0.95$.
}
\end{center}
\end{figure}

Fig.~\ref{critico} shows the estimation of the critical point, along with the $\nu$
critical exponent. From the fit we obtain the values $T_c=1.2180(29)$ and $\nu=0.65(3)$, which are in good agreement with
the reported values for the SW with $\lambda=1.5$ and the critical exponent for the Ising model,
$\nu=0.6302(1)$~\cite{Campbell2011}, respectively. The value for the non universal parameter $a$ obtained from the same fit is $a=0.95(8)$. In table~\ref{finales} we are summarizing our results, along with
previous reported values of $T^*_c$ and $\nu$.
\begin{center}
\begin{table}[ht!]
\caption{\label{finales}
Critical parameters for the SW fluid with $\lambda=1.5$ obtained in this work and those from literature.
}
\begin{tabular}{ccccc}
\hline
\hline
$T^*_c$ &  ~ & $\nu$ & ~ & Source\\
\hline
  1.2180(29) & ~ & 0.65(3) & ~ &  This work \\
  1.2179(3) & ~ & 0.63(3) & ~ & Orkoulas {\em et al.}~\cite{Orkoulas2001} \\
  1.218 & ~ & $-$ & ~ & Del R\'io {\em et al.}~\cite{Delrio2002} \\
  1.2172(7) & ~ & $-$ & ~ & Singh {\em et al.}~\cite{Singh2003} \\
\hline
\hline
\end{tabular}
\end{table}
\end{center}
The incertitude in the critical point found in this work is one order of
magnitude bigger that the reported in Refs.~\cite{Orkoulas2001,Singh2003}, this is due to
the low number of independent simulations performed. The issue will be addressed
in future works.

\section{Conclusions}

\label{conclusiones}

We present a novel methodology for the evaluation of the critical temperature and the correlation length
critical exponent for a SW fluid with interaction
range $\lambda=1.5$. The method is able to obtain reliable results, even with small statistic, without
previous assumptions about the value of $T^*_c$ and $\nu$. In future works we will be exploring the possibility to
implement the method in systems with bigger asymmetry, in particular for another ranges in the SW fluid or the
Lennard-Jones fluid. Another problem that can be addressed is the complete evaluation of the VLE curve using the
$\mu^*$ versus $\rho^*$ curves. From the coexistence curves it will be possible to evaluate the critical density
using the Wegner expansion~\cite{Wegner1972}. We must point out that
our method fails at high densities, in the same way that the Widom
test-particle-insertion (TPI) method~\cite{Widom1963} fails, and it is not possible to obtain numerical values 
of the chemical potential for the densities reported
by Lab\'ik {\em et al.}~\cite{Labik1999}. In a possible future work we will try to implement a variation of
the scaled-particle Monte Carlo (SP-MC) method to our algorithm for high densities.

\section*{Acknowledgments}
The author thanks Ana Laura Benavides and Alejandro Gil-Villegas for useful comments and for the critical reading of the manuscript. 
This research was supported by Universidad de Guanajuato (M\'exico) under Proyecto DAIP 879/2016.

\clearpage
%\begin{thebibliography}{00}

\bibliography{bibliografia}

%\end{thebibliography}
\end{document}